\documentclass[superscriptaddress,showpacs,prl,preprint,onecolumn]{revtex4}%
\usepackage{graphicx}
\usepackage{amsmath}
\usepackage{amsfonts}
\usepackage{amssymb}%
\begin{document}
\author{Gerrit E. W. Bauer}
\affiliation{Department of NanoScience, Delft University of Technology, 2628 CJ Delft, The Netherlands}
\author{Arne Brataas}
\affiliation{Department of Physics, Norwegian University of Science and Technology, N-7491
Trondheim, Norway}
\author{Yaroslav Tserkovnyak}
\affiliation{Lyman Laboratory of Physics, Harvard University, Cambridge, Massachusetts
02138, USA}
\author{Bertrand I. Halperin }
\affiliation{Lyman Laboratory of Physics, Harvard University, Cambridge, Massachusetts
02138, USA}
\author{Maciej Zwierzycki}
\affiliation{Faculty of Applied Physics and MESA Research Institute, University of Twente,
7500 AE Enschede, The Netherlands}
\author{Paul J. Kelly}
\affiliation{Faculty of Applied Physics and MESA Research Institute, University of Twente,
7500 AE Enschede, The Netherlands}
\title{Dynamic ferromagnetic proximity effect in photoexcited semiconductors}

\begin{abstract}
The spin dynamics of photoexcited carriers in semiconductors in contact with a
ferromagnet is treated theoretically and compared with time-dependent Faraday
rotation experiments. The long time response of the system is found to be
governed by the first tens of picoseconds in which the excited plasma
interacts strongly with the intrinsic interface between semiconductor and
ferromagnet in spite of the existence of a Schottky barrier in equilibrium.

\end{abstract}
\pacs{72.25.Mk,78.47.+p,75.70.-i,76.70.Fz}
\date{\today}
\maketitle

Magnetoelectronics, \textit{i.e}. the science and technology of using
ferromagnets in electronic circuits, is divided into two subfields,
metal-based \cite{GMR} and semiconductor-based magnetoelectronics
\cite{Optics}, with little common ground.

Metal researchers focus mainly on topics derived from giant magnetoresistance,
the large difference in the DC conductance for parallel and antiparallel
magnetization configurations of magnetic multilayers, where theoretical
understanding has progressed to the stage of materials-specific predictions
\cite{SSP}. More recently, the dynamics of the magnetization vectors in the
presence of charge and spin currents has received a lot of attention
\cite{Slon96,Tsoi98,Heinrich,Ralph}.

Semiconductor-based magnetoelectronics is motivated by the prospect of
integrating new functionalities with conventional semiconductor electronics.
The emphasis has been on the basic problem of spin injection into
semiconductors, theoretical understanding is less advanced and detailed
electronic structure calculations are just starting \cite{Wunnicke,Zwierzycki}%
. Unlike metals, semiconductors can be studied by optical spectroscopies such
as the powerful time-resolved Faraday or Kerr rotation techniques, in which a
selected component of a spin-polarized excitation cloud in the semiconductor
can be monitored on $%
\operatorname{ps}%
$ time scales \cite{Awschalom}. In \textit{n}-doped GaAs, these experiments
revealed long spin-coherence of the order of $%
\operatorname{\mu s}%
$ \cite{Awschalom}. When the semiconductor is in contact with a ferromagnet,
an initially unpolarized electron distribution prepared using linearly
polarized light, was found to very quickly acquire a spin polarization - the
dynamic ferromagnetic proximity (DFP) effect \cite{Kawakami,Epstein1,Epstein2}%
. In turn DFP efficiently imprints spin information from a ferromagnet onto
nuclear spins by dynamic nuclear polarization, opening new options for quantum
information storage.

In this Letter we wish to show that metal and semiconductor-based
magnetoelectronics can both be understood in terms of coherent spin
accumulations. Specifically, the DFP can be treated by the same formalism that
successfully describes the dynamics of the magnetization vector in metallic
hybrids \cite{Gilbert,Heinrich}. The experiments can be understood in terms of
a time-dependent interaction between the conduction-band electrons and the
ferromagnet in a \textquotedblleft fireball-afterglow\textquotedblright%
\ scenario. Photoexcited holes are instrumental in helping the electrons to
overcome the Schottky barrier between metal and semiconductor in the first
$\lesssim50%
\operatorname{ps}%
$ (\textquotedblleft fireball\textquotedblright\ regime) and induce the
proximity effect. The interaction weakens with vanishing hole density, thus
preventing fast decay of the created spin accumulation in the
\textquotedblleft afterglow\textquotedblright.

Two groups have already contributed important insights into this problem.
Ciuti \textit{et al.} \cite{Sham} interpreted the DFP in terms of a
spin-dependent reflection of electrons at a ferromagnetic interface through a
Schottky barrier in equilibrium, but did not address the time dependence of
the problem. Gridnev \cite{Gridnev} did investigate the dynamics of the
photoexcited carriers, but postulated a phenomenological relaxation tensor
with a specific anisotropy that we find difficult to justify. We show
how\ both approaches can be unified and extended by \textit{ab initio}
magnetoelectronic circuit theory \cite{Brataas00,Bauer03,Xia01,Xia02}.

We first summarize the experimental evidence \cite{Epstein1,Epstein2}.
Initially, a $\sim100$ $%
\operatorname{fs}%
$ pulse with frequency close to the band gap is absorbed by the semiconductor
(100 $%
\operatorname{nm}%
$ of GaAs). The polarization state is then monitored by time-dependent Faraday
rotation measurements of the coherent spin precession in an applied magnetic
field. The homogeneously excited carriers (\textquotedblleft
fireball\textquotedblright) thermalize within a $%
\operatorname{ps}%
$, in which the holes also lose any initial spin polarization. The interaction
time-scale with the ferromagnet (Fe or MnAs) can be deduced from the rise time
of the polarization after excitation with a linearly polarized (LP) light
pulse to be $\leq50$ $%
\operatorname{ps}%
$ \cite{Epstein1}. For long delay times (\textquotedblleft
afterglow\textquotedblright), the spin relaxation is very slow ($>2$ $%
\operatorname{ns}%
$), comparable to GaAs reference samples in the absence of a ferromagnet. The
sample can be also excited by circularly polarized light (CP), in which case
the fireball is polarized from the outset. Dynamic nuclear polarization (DNP)
by the hyperfine interaction can be detected by deviations of the precession
frequency from the bare Larmor frequency, \textit{i.e}. a modified g-factor
\cite{Kawakami}. DNP should vanish when the external magnetic field is normal
to the spin accumulation. It therefore remains to be explained that Epstein
\textit{et al.} \cite{Epstein1} observe a modified $g-$factor for this
configuration that differs for LP and CP excitation. Interesting additional
information relates to the material dependence, indicating that the
polarization induced by Fe is of opposite sign to that induced by MnAs
\cite{Epstein1}, and to the modulation of the afterglow Larmor frequency by an
applied bias \cite{Epstein2}.

Let us consider a semiconductor (Sc) film in which a non-equilibrium electron
chemical potential $\langle\mu|=\langle\mu_{c},\vec{\mu}_{s}|$ is excited with
charge $\mu_{c}$ and spin $\langle\vec{\mu}_{s}|\equiv\langle\mu_{x},\mu
_{y},\mu_{z}|$ accumulation (in energy units). The bilayer parallel to the
$\left(  yz\right)  $ plane consists of a semiconductor in contact with a
metallic ferromagnetic film (F) with fixed single-domain magnetization in the
direction of the unit vector $\vec{m}$. By its relatively huge density of
states a metallic ferromagnet may be treated as a reservoir in equilibrium. A
charge $\left(  I_{c}\right)  $ and spin $\left(  \vec{I}_{s}\right)  $
current $\langle I|=\langle I_{c},\vec{I}_{s}|$ (in units of reciprocal time)
flows through the ferromagnet/semiconductor (F%
$\vert$%
Sc) interface, which is governed by the spin-dependent (dimensionless)
conductances $g_{\uparrow\uparrow}$ and $g_{\downarrow\downarrow}$ as well as
the complex spin-mixing conductance $g_{\uparrow\downarrow}$ \cite{Brataas00}%
$.$ Physically, the real part of the mixing conductance expresses the angular
momentum transfer to and from the ferromagnet, such as the strength of the
spin-current induced magnetization torque \cite{Slon96,Bauer03} or non-local
Gilbert damping \cite{Gilbert}, whereas the imaginary part is an effective
magnetic field \cite{Huertas00,Stiles,Huertas02}. The microscopic expression
for the conductances is Landauer-like\
\begin{equation}
g_{\sigma\sigma^{\prime}}=\sum_{nm}[\delta_{nm}-r_{nm}^{\sigma}(r_{nm}%
^{\sigma^{\prime}})^{\ast}], \label{mix}%
\end{equation}
where the reflection coefficient $r_{nm}^{\sigma}$ of an electron in the Sc
with spin $\sigma$ at the Sc%
$\vert$%
F contact between $n-$th and $m-$th transverse modes is accessible to
\textit{ab initio} calculations \cite{Xia01,Xia02,Wunnicke,Zwierzycki}. The
time dependence of the system is governed by charge and spin conservation
\cite{Brataas00}:%
\begin{align}
-2h\mathcal{D}\left(  \frac{d\mu_{c}}{dt}\right)  _{bias}  &  =\left(
g_{\uparrow\uparrow}+g_{\downarrow\downarrow}\right)  \left(  \mu_{c}%
-e\varphi\right) \nonumber\\
&  +\left(  g_{\uparrow\uparrow}-g_{\downarrow\downarrow}\right)  \left(
\vec{m}\cdot\vec{\mu}_{s}\right)  \label{charge}%
\end{align}%
\begin{gather}
-2h\mathcal{D}\left(  \frac{d\vec{\mu}_{s}}{dt}\right)  _{bias}%
=2\operatorname{Re}g_{\uparrow\downarrow}\vec{\mu}_{s}+2\operatorname{Im}%
g_{\uparrow\downarrow}\left(  \vec{m}\times\vec{\mu}_{s}\right) \nonumber\\
+\left[  \left(  g_{\uparrow\uparrow}-g_{\downarrow\downarrow}\right)  \left(
\mu_{c}-e\varphi\right)  +\left(  g_{\uparrow\uparrow}+g_{\downarrow
\downarrow}-2\operatorname{Re}g_{\uparrow\downarrow}\right)  \left(  \vec
{m}\cdot\vec{\mu}_{s}\right)  \right]  \vec{m} \label{spin}%
\end{gather}
where $\mathcal{D}$ is the Sc single-spin energy density of states. The
electrostatic potential $\varphi$ due to an applied bias and/or a charge
imbalance between electrons and holes will be disregarded in the following
(see below). In the presence of a magnetic field $\vec{B}$, the sum of
externally applied and hyperfine (Overhauser) fields with ordered nuclear
spins, we have to add
\begin{equation}
\left(  \frac{d\vec{\mu}_{s}}{dt}\right)  _{field}=-\frac{g_{e}\mu_{B}}{\hbar
}\vec{B}\times\vec{\mu}_{s}%
\end{equation}
where $g_{e}$ is the electron g-factor ($\approx-0.4$ in GaAs) and $\mu_{B}$
the Bohr magneton. These equations can be summarized in terms of a $4\times4$
matrix equation:%
\begin{equation}
-T_{I}\frac{d|\mu\rangle}{dt}=\mathbf{\Gamma}|\mu\rangle. \label{Kinetic}%
\end{equation}
where $T_{I}=2h\mathcal{D}/g$ is an interface-mediated relaxation time in
terms of the total conductance $g=g_{\uparrow\uparrow}+g_{\downarrow
\downarrow}$. Choosing $\vec{m}$\textbf{\ }parallel to the $z-$axis:%
\begin{equation}
\mathbf{\Gamma}=\left(
\begin{array}
[c]{cccc}%
1 & 0 & 0 & p\\
0 & \eta_{r} & -\eta_{i}-\Omega_{z} & \Omega_{y}\\
0 & \eta_{i}+\Omega_{z} & \eta_{r} & -\Omega_{x}\\
p & -\Omega_{y} & \Omega_{x} & 1
\end{array}
\right)  ,
\end{equation}
$\left\vert \Omega_{\alpha}\right\vert /T_{I}=\left\vert g_{e}\right\vert
\mu_{B}B_{\alpha}/\hbar$ is the Larmor frequency, the mixing conductance has
been normalized as $\eta=2g_{\uparrow\downarrow}/g$ with subscripts $i$ and
$r$ denoting its real and imaginary part, and the polarization is defined as
$p=\left(  g_{\uparrow\uparrow}-g_{\downarrow\downarrow}\right)  /g$. Eq.
$\left(  \text{\ref{Kinetic}}\right)  $ can be solved easily \cite{Gridnev}
for the boundary conditions corresponding to LP excitation $\langle\mu
^{LP}\left(  0\right)  |=\langle1,0,0,0|$ or circularly polarized (CP)
excitation with wave vector in the $x-$direction $\langle\mu^{CP}\left(
0\right)  |=\langle1,1,0,0|$.

When the conductance is expressed in terms of an interface transparency
parameter $\kappa$ times the intrinsic Sc single-spin Sharvin conductance, we
can write:
\begin{equation}
T_{I}\approx\frac{3.5}{\kappa}\frac{L}{100\text{ }\mathrm{nm}}\sqrt
{\frac{m^{\ast}}{0.067m_{0}}\frac{10\text{ }\mathrm{meV}}{\varepsilon}}%
\operatorname{ps}%
. \label{relax}%
\end{equation}
For an Ohmic interface with $\kappa\simeq0.1,$ $L_{Sc}=100$ \textrm{$%
\operatorname{nm}%
$} and at a characteristic electron kinetic energy (depending on doping and
excitation density) of $\varepsilon=10$ $\mathrm{meV,}$ the time constant for
GaAs (effective mass $m^{\ast}=0.067m_{0}$) is $T_{I}\sim35$ \textrm{$%
\operatorname{ps}%
$}$,$ of the order of the experimental rise time of the proximity effect
\cite{Epstein1}. At long time scales, experiments find $\breve{T}_{I}>2$
$\mathrm{%
\operatorname{ns}%
}$, which corresponds to a strongly reduced transparency of $\breve{\kappa
}\lesssim0.002.$

With a few exceptions (notably InAs), Schottky barriers are formed at
metal-semiconductor interfaces when interface states in the gap of the
semiconductor become filled giving rise to space charges. Photoexcited holes
are strongly attracted by the barrier, thereby dragging the electrons with
them \cite{Flatte} and/or screen the barrier. The observed large $\kappa$ in
the \textquotedblleft fireball\textquotedblright\ regime reflects the
facilitation of electron transport to the Sc%
$\vert$%
F interface by the holes. In the limit of predominant ambipolar electron-hole
transport (and absence of an applied bias) we can justify negelect of
electrostatic potential $\varphi,$ but the electron conductance $g$ might be
affected by the scattering of the holes.

At long time scales, the holes disappear into the ferromagnet or recombination
with electrons in the semiconductor, but the electron spin accumulation
persists \cite{Sham}. In this afterglow, remaining space charges vanish when
the sample is grounded and net charge transport is suppressed. Eq.
(\ref{charge}) thus vanishes and
\begin{equation}
-\breve{T}_{I}\frac{d\vec{\mu}_{s}}{dt}=\mathbf{\breve{\Gamma}}_{s}\vec{\mu
}_{s}. \label{Bloch}%
\end{equation}
with%
\begin{equation}
\mathbf{\breve{\Gamma}}_{s}=\left(
\begin{array}
[c]{ccc}%
\breve{\eta}_{r} & -\breve{\eta}_{i}-\breve{\Omega}_{z} & \breve{\Omega}_{y}\\
\breve{\eta}_{i}+\breve{\Omega}_{z} & \breve{\eta}_{r} & -\breve{\Omega}_{x}\\
-\breve{\Omega}_{y} & \breve{\Omega}_{x} & 1-\breve{p}^{2}%
\end{array}
\right)  .
\end{equation}
where the decoration $\smallsmile$ indicates electron transport through the
full Schottky barrier.

The kinetic equations are valid when the system is diffuse or chaotic (as a
result of interface roughness or bulk disorder). The ferromagnetic elements
should have an exchange splitting $\Delta$ which is large enough that the
magnetic coherence length $\ell_{c}=\hbar/\sqrt{2m\Delta}<\min\left(
\ell,L_{F}\right)  ,$ where $L_{F}$\ is the thickness of the ferromagnetic
layer (typically $50$ $%
\operatorname{nm}%
$). These conditions are usually fulfilled in hybrid systems except for very
thin layers with nearly perfect interfaces. The spin relaxation time in GaAs
is taken to be very long. We also require that $L_{Sc}<\sqrt{2\hbar v_{F}%
\ell/3g_{e}\mu_{B}B_{ext}}$ for diffuse systems \cite{Huertas00} or
$L_{Sc}<\hbar v_{F}/g_{e}\mu_{B}B_{ext}$ for ballistic systems, where $\ell$
is the mean free path, $v_{F}$ the Fermi velocity and $B_{ext}$ the externally
applied field. For samples with $L_{Sc}=100~%
\operatorname{nm}%
$ the applied magnetic fields should therefore not much exceed 5 T.

Thus a microscopic justification can be given for Gridnev's phenomenological
relaxation tensor \cite{Gridnev}. But whereas we can explain that the
longitudinal relaxation $\sim g\ $or $\sim\breve{g}\left(  1-\breve{p}%
^{2}\right)  $ can differ from the transverse components $\left(
\sim\operatorname{Re}g_{\uparrow\downarrow}\text{ or }\sim\operatorname{Re}%
\breve{g}_{\uparrow\downarrow}\right)  ,$ Gridnev postulated a large
difference between the two components \emph{normal} to $\vec{m}.$ Such large
magnetic anisotropies can be excluded for Fe and the scenario sketched by
Gridnev can not be \ a generic explanation for all experiments. To explain the
proximity polarization, Gridnev's $3\times3$ Bloch equation \emph{must} be
extended to the $4\times4$ kinetic equation (\ref{Kinetic}) that includes a
charge current component.

First-principles calculations of the bare interface conductance without
Schottky barrier provide a first indication of the transport properties in the
first tens of $%
\operatorname{ps}%
$. We choose here the Fe%
$\vert$%
InAs system, which apart from the Schottky barrier, is very similar to Fe%
$\vert$%
GaAs, and as such, of great interest in itself. Table 1 summarizes results
obtained by scattering matrix calculations with a first-principles
tight-binding basis described in Ref. \cite{Zwierzycki}. We find the reversal
of polarization sign with disorder as noted before \cite{Zwierzycki}, which
may explain the negative polarization found for Fe%
$\vert$%
GaAs \cite{Epstein1}. The real part of the mixing conductance $\eta_{r}$
(torque) is close to unity similar to metallic interfaces, but in contrast to
these, the imaginary part $\eta_{i}$ is strongly enhanced$.$ The latter can be
explained by the focus on a small number of states with wave vectors close to
the origin, which prevents the averaging to zero found in metals
\cite{Xia02}.\begin{table}[ptb]
\begin{center}
\begin{ruledtabular}
\begin{tabular}
[c]{ccccccc}
&           & $G\left[  1/f\Omega m^{2}\right]  $ & $p$ & $\kappa$ &
$\operatorname{Re}\eta$ & $\operatorname{Im}\eta
$                         \\ \hline\hline Clean & Fe$|$InAs & $1.5\cdot
10^{-5}$ & $0.98$ & $0.14$ & $1.3$ & $-1.3$  \\
& Fe$|$AsIn & $3.6\cdot10^{-5}$ & $0.88$ & $0.35$ & $1.6$ & $-1.05$ \\ \hline
Dirty & Fe$|$InAs & $5.7\cdot10^{-5}$ & $-0.29$ & $0.56$ & $1.1$ & $-0.18$\\
& Fe$|$AsIn & $7.4\cdot10^{-5}$ & $-0.22$ & $0.71$ & $1.3$ & $-0.30$      \\
\end{tabular}
\end{ruledtabular}
\end{center}
\caption{\textit{Ab initio} interface transport parameters for a clean
InAs$|$Fe (001) interface with In (or As) termination. $G$ is the electric
conductance $e^{2}g/h,$ $p$ the polarization, $\kappa$ the ratio between $G$
and the intrinsic Sc Sharvin conductance $\left(  5.2\cdot10^{-5}/\left(
f\Omega m^{2}\right)  \right)  $, $\eta$ the relative mixing conductance, all
at a kinetic energy of 20 meV in the InAs conduction band. The dirty
interfaces are modeled as a monolayer of random alloy with 1/4 of the
interface In (or As) replaced by Fe in a 7$\times$7 lateral supercell.}%
\end{table}

We now model the experiments of Epstein \textit{et al.} on GaAs%
$\vert$%
MnAs (concentrating on Fig. 1 in \cite{Epstein1}) in which an external
magnetic field of 0.12 T lies in the interface $(yz)$ plane. Analytic
solutions of Eq. (\ref{Kinetic}) can be obtained for $\eta_{r}=1\left(
=\breve{\eta}_{r}\right)  \ $(see Table 1 and \cite{Xia02}) such that all
modes are exponentially damped by a single interface relaxation time $T_{I}.$
Time is measured in units of $T_{I}$ and the polarization is chosen to be
$p=1\left(  =\breve{p}\right)  $. In the fireball regime the quality factor
$\Omega_{y}\ \ll1,$ and we adopt $\eta_{i}=-1\left(  =\breve{\eta}_{i}\right)
$. For LP excitation the charge component relaxes in favor of the
$z-$component polarized along the magnetization direction, $2\mu_{z}\left(
t\right)  =e^{-\left(  1-p\right)  t/T_{I}}-e^{-\left(  1+p\right)  t/T_{I}}$
for $\Omega_{y}=0$, which is the essence of the DFP effect. The time scale on
which the Schottky barrier recovers determines (together with $p$) the modulus
of the spin accumulation in the afterglow. It is of the same order, but
smaller, than $T_{I}$ because of competing electron-hole recombination in the
semiconductor. Our Fig. 1 is similar to Fig. 1 in \cite{Epstein1}, but
additional experimental data on a short time scale are required to guide the
development\ of a more refined model. \begin{figure}[ptb]
\begin{center}
\includegraphics[
height=7.1 cm,
width=5.3 cm
]{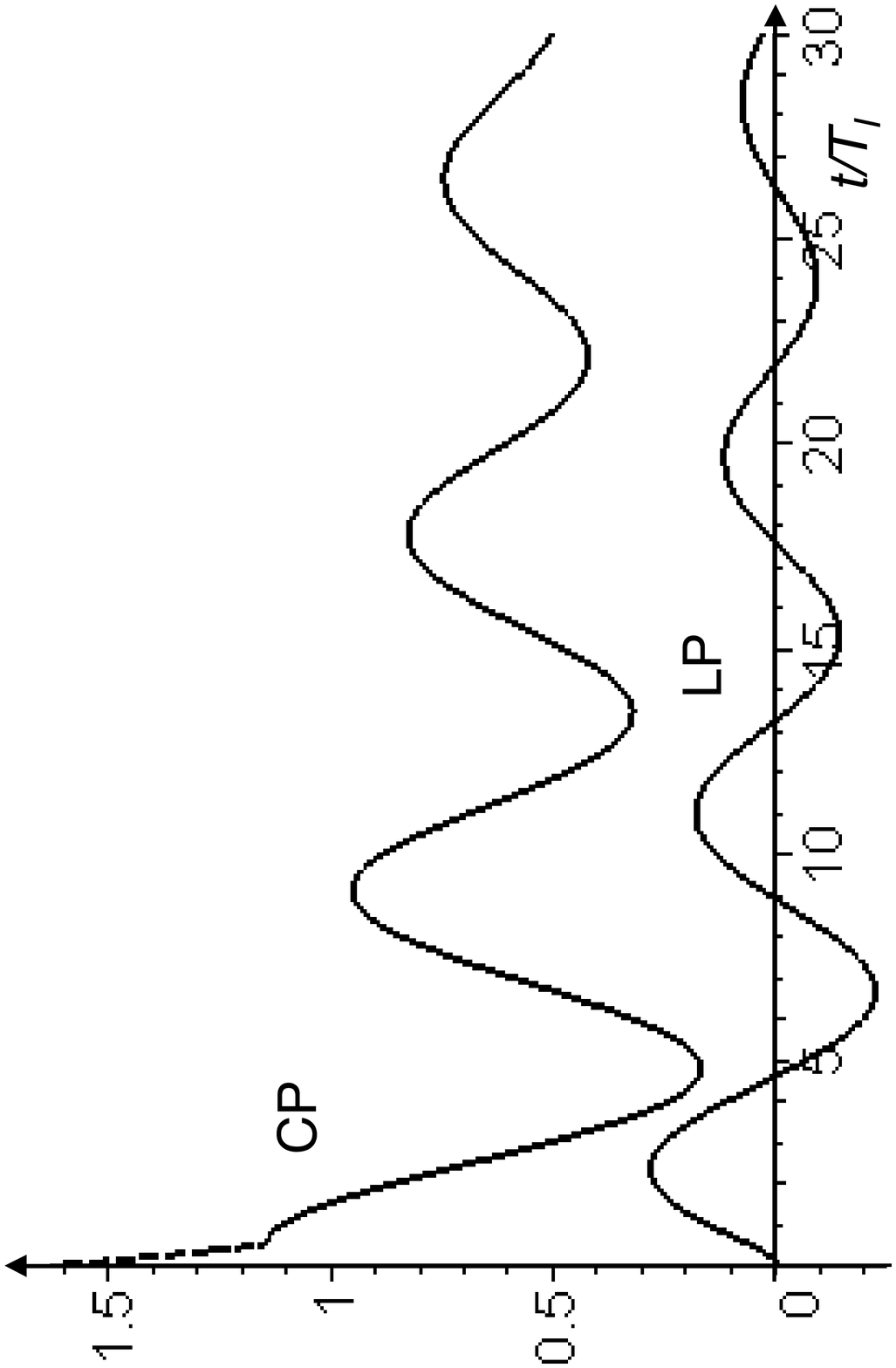}
\end{center}
\caption{Spin-dynamics in the\ "fireball-afterglow" scenario of an excited
semiconductor in proximity with a ferromagnet polarized in the $z-$direction.
A (DNP enhanced) magnetic field 0.24 T is applied in the $y-$direction.
Plotted is $\mu_{x}\left(  t\right)  $ in arbitrary units for CP (shifted
upwards by 0.6) and LP excitation, with wave vector in $x-$direction. Time is
measured in units of $T_{I}$ $.$ The transition from fireball to afterglow
with $\breve{T}_{I}=10T_{I}$ is taken to be abrupt at $t=T_{I}/2.$}%
\label{Simulation}%
\end{figure}As mentioned above, a modified Larmor frequency has been observed
\cite{Epstein1} even when the photon wave vector is normal to the field. This
is at odds with the notion that a CP excited spin accumulation should rotate
around the field without net angular momentum transfer to the nuclei. This
\textit{could} be evidence for a DFP effect for the CP configuration. In the
fireball, a significant $\eta_{i}$ acts like a magnetic field in the
$z-$direction, causing the initial spin ensemble to precess into the direction
of the external magnetic field, which is then able to polarize the nuclear
spins. This effect is weaker for LP excitation since, in the brief fireball
interval, any spin accumulation has to be generated before it can precess.

For LP excitation the Larmor frequency depends \cite{Epstein2} on an applied
bias, proving that the electrostatic potential $\varphi$ can, in general, not
be neglected in Eqs. (2,3). This does not invalidate our qualitative arguments
since, compared to the bare interface exchange the modifications needed to
explain the shifts in the effective Larmor frequencies are small, beyond the
accuracy of our model. The decreasing spin lifetime in the afterglow with
increasing bias has been explained by inhomogeneus nuclear polarization
\cite{Epstein2}, but lowering the Schottky barrier by a forward bias also
reduces $\breve{T}_{I}$.

In conclusion, we propose a physical picture for the spin dynamics of
photoexcited carriers in semiconductor%
$\vert$%
ferromagnet bilayers. The experiments can be understood in terms of at least
two time scales. In the first 50 ps or so, the photoexcited carriers screen
the Schottky barrier efficiently and the interaction of the electrons with the
ferromagnet is described by nearly intrinsic interface conductances that can
be calculated from first principles. After delay times of
$>$%
100 ps, the Schottky barrier protects the semiconductor carriers from fast
decay and any residual exchange interaction is very weak. More insight into
the interaction of carriers in semiconductors with ferromagnets could be
gained by a faster ($%
\operatorname{ps}%
$) time resolution and higher applied magnetic fields. Quantitative
explanation of the experiments requires self-consistent modelling of the
combined electron and hole carrier dynamics as well as \textit{ab initio}
calculations of the interface scattering matrices for electrons and holes.

We thank D.D. Awschalom, Y. Kato, and R.J. Epstein for helpful discussions on
the Faraday rotation experiments. This work has been supported by the FOM, the
NEDO joint research program \textquotedblleft Nano
Magnetoelectronics\textquotedblright, DARPA award MDA 972-01-1-0024, and the
European Commission's RT Network \emph{Computational Magnetoelectronics}
(Contract No. HPRN-CT-2000-00143).

\end{document}